\documentclass[fleqn,usenatbib]{mnras}

\usepackage[usenames,dvipsnames]{xcolor}
\usepackage{textfit}
\usepackage{float}
\usepackage[T1]{fontenc}
\usepackage{type1cm} 

\usepackage[flushleft]{threeparttable}

\DeclareRobustCommand{\VAN}[3]{#2}
\let\VANthebibliography\thebibliography
\def\thebibliography{\DeclareRobustCommand{\VAN}[3]{##3}\VANthebibliography}

\usepackage{graphicx}	% Including figure files
\usepackage{amsmath}	% Advanced maths commands
\usepackage{soul}
\title[Vegetation and habitability]{Impact of vegetation albedo on the habitability of Earth-like exoplanets}

\author[E. Bisesi et al.]{
E. Bisesi $^{1,2}$\thanks{E-mail: erica.bisesi@inaf.it},
G. Murante $^{1,2,3,4}$,
A. Provenzale $^{2}$,
L. Biasiotti $^{1,5}$,
J. von Hardenberg $^{6,7}$,
S. Ivanovski $^{1,5}$,
\newauthor
M. Maris $^{1,3,4}$,
S. Monai $^{1}$,
L. Silva $^{1,3}$,
P. Simonetti $^{1}$,
G. Vladilo $^{1}$
\\
$^{1}$INAF $-$ Astronomical Observatory of Trieste, Via G. Tiepolo 11, 34143 Trieste, Italy\\
$^{2}$CNR $-$ Institute of Geosciences and Earth Resources, via G. Moruzzi 1, 56124 Pisa, Italy\\
$^{3}$Institute for Fundamental Physics of the Universe, Via Beirut 2, 34151 Trieste, Italy\\
$^{4}$ICSC $-$ National Research Center in High Performance Computing, Big Data e Quantum Computing, \\ Via Magnanelli 2, 40033 Casalecchio di Reno (BO), Italy
\\
$^{5}$University of Trieste, Piazzale Europa 1, 34127 Trieste, Italy\\
$^{6}$Polytechnic University of Turin $-$ DIATI, Corso Duca degli Abruzzi 24, 10129 Turin, Italy\\
$^{7}$CNR $-$ Institute of Atmospheric Science and Climate, Corso Fiume 4, 10133 Turin, Italy\\
}

\date{Accepted XXX. Received YYY; in original form ZZZ}

\pubyear{2023}

\begin{document}
\label{firstpage}
\pagerange{\pageref{firstpage}--\pageref{lastpage}}
\maketitle

\begin{abstract}
%%%%%%%%%%%%%%%%%%%%%%%%%
%The abstract should briefly describe the aims, methods, and main results of the paper.
%It should be a single paragraph not more than 250 words (200 words for Letters).
%No references should appear in the abstract.
Vegetation can modify the planetary surface albedo via the Charney mechanism, as plants are usually darker than the bare surface of the continents. We updated ESTM (\textit{Earth-like Surface Temperature Model}) to incorporate the presence, distribution and evolution of two dynamically competing vegetation types that resemble grasslands and trees (the latter in the double stages of life: adults and seedlings). The newly developed model was applied to estimate how the climate-vegetation system reaches equilibrium across different rocky planetary configurations, and to assess its impact on temperature and habitability. With respect to a world with bare granite continents, the effect of vegetation-albedo feedback is to increase the average surface temperature. Since grasses and trees exhibit different albedos, they affect temperature to different degrees. The ultimate impact on climate depends on the outcome of the competition between these vegetation types. 
The change in albedo due to vegetation extends the habitable zone and enhances the overall planetary habitability beyond its traditional outer edge. This effect is especially relevant for planets that have a larger extension of continents than Earth. For  Earth, the semi-major axis \textit{d} = 1.04 UA represents the turning point where vegetation enhances habitability from \textit{h} = 0.0 to \textit{h} = 0.485 (in the grass-dominance case), to \textit{h} = 0.584 (in the case of coexistence between grasses and trees), and to \textit{h} = 0.612 (in the tree-dominance case). This illustrates the transition from a snowball state to a planet with intermediate habitability at the outer edge of the circumstellar habitability zone. 
%%%%%%%%%%%%%%%%%%%%%%%%%
\end{abstract}


%%%%%%%%%%%%%%%%% BODY OF PAPER %%%%%%%%%%%%%%%%%%

\section{Introduction}
\label{sec:introduction}

The study of exoplanetary climates plays a crucial role in Astrobiology, the science that investigates the origin, evolution and distribution of life in the Universe (see e.g. \citep{Williams1997a} and references therein). The climate of a rocky planet is of paramount importance to establish the possible presence of liquid water at its surface \citep{Dole1964, Hart1979, Kasting1993}, and thus, in the wide sense generally accepted by the scientific community, its habitability. Climate depends on a great number of factors and parameters, both astrophysical (e.g., luminosity of the central star(s), distance of the planet to the star, orbital eccentricity) and planetary (inclination of the rotation axis, day duration, atmospheric composition and pressure, presence and fraction of oceans, presence and distribution of continental masses, type of soil and orography, etc.). Some of these parameters have been or will be measured in the near future; data on the others will not be available in short time, and maybe not even in the far future \citep{Li2022}.

\indent 
Given the level of uncertainty on many of such planetary characteristics, simple conceptual climate models lend themselves to better serving at parameter space exploration than the more complex Global Climate Models (GCM; see e.g. \citep{Provenzale2014} for a simple discussion of the hierarchy of climate models), for example in reason of their much lower computational cost. “Simple” models allow running thousands of numerical experiments at the same computational time of a single GCM’s run. Within the broad family of EBMs (Energy Balance Models), a representative example of such a streamlined model is ESTM (Earth-like Surface Temperature Model: \citep{Vladilo2013, Vladilo2015} $-$ V13, V15 from now on, \citep{Biasiotti2022}). As all EBMs, ESTM is a type of climate tool commonly employed in studies of the so-called “habitable zone” \citep{Kasting1988, Kasting1993}, and is based on the numerical solution of a modified diffusion equation for the meridional heat transfer, coupled with a radiative-convective  atmospheric column model to account for the vertical transport of radiation (downward and upward). This model has been used for studying a number of characteristics of possible exoplanets candidates (see e.g. \citep{Silva2017, Murante2020}, and references therein). It has also been used to explore how the outer edge of the habitable zone varies by modifying basic planetary parameters \citep{Biasiotti2022}.

\indent 
On Earth, a well-known regulator of planetary climate is its vegetation cover. Vegetation can modify the planetary surface albedo, as it is usually darker than the bare surface of the continents (Charney mechanism: \citep{Charney1975}; see also \citep{Baudena2009} $-$ B09 from now on, \citep{CrestoAleina2013}, and references therein). Other relevant vegetation-climate interactions include the effects on the “fast” and “slow” carbon cycles, the former through the way in which plants, or more generally the biosphere, can absorb (by photosynthesis) and emit (by respiration) carbon into the atmosphere, and the second relating to the way terrestrial vegetation can alter the speed of rock weathering and thus the pace of the geological carbon cycle. Both processes change the carbon dioxide or methane fraction in the atmosphere and, as a consequence, the global greenhouse effect. Vegetation evapotraspiration is also important, both in reason of its direct surface cooling effect and of the emission of water vapour that can participate in the hydrological cycle \citep{CrestoAleina2013, Trostl2016, Zhao2017}.

\indent
While a complete understanding of the effects of vegetation would require consideration of all three effects, in this work we focus on the albedo effect of vegetation, and adopt the ESTM approach to investigate the relevance of the Charney mechanism on exoplanet's habitability. A central aspect of our exploration is the inclusion of vegetation competition and coexistence, considering more than one type of vegetation – namely, tree-like and grass-like plants. We thus introduce an updated version of ESTM, which takes into account the presence, distribution and temporal dynamics of different vegetation types. Differential equations describing how changes in the vegetation distribution impact changes in temperature via the albedo feedback, and vice-versa, have been implemented into the model. The newly developed version of ESTM, which includes vegetation dynamics, has been applied to estimate (a) how different and dynamically competing vegetation types (inspired by terrestrial trees and grasses) reach an equilibrium distribution on a planet depending on its main properties (the most straightforward of these being the insolation, i.e., a combination of stellar luminosity and planet distance from the star), and (b) how the presence of vegetation impacts the planet's surface temperature and habitability, thus modifying the corresponding circumstellar habitable zone.

\section{Scientific questions}
\label{sec:scientquest}

The aim of this study is to analytically and numerically combine ESTM (V13, V15) with an idealized model for tree–grass dynamics such as that described in B09, to estimate how dynamically competing vegetation types impact  the surface albedo, and consequently the temperature and habitability of rocky, terrestrial-type exoplanets. This endeavor should be seen as a first step of a research program aimed at including the main climate-vegetation feedbacks known for Earth in exoplanetary habitability assessments.\\ 

\indent 
On Earth, vegetation growth depends on several environmental factors, including temperature and soil moisture. Given the structure of ESTM, in keeping with previous works \citep{Charney1975, vonBloh1997} we relate vegetation growth and mortality to the surface temperature, which is well handled by ESTM. The use of temperature as a simplified proxy for the intensity of the water cycle was often used in conceptual models of climate-biosphere interactions, see e.g. \citep{vonBloh1997}, given the difficulties in properly handling a full hydrological cycle in simplified energy-balance models.\\
\indent
The surface temperature of a planet depends, in turn, on a number of stellar, orbital and geographic parameters. For exploring the role of vegetation, we investigated a set of standard configurations in parameter space: (i) the Earth as a reference benchmark, (ii) a pseudo-Earth with its rotation axis perpendicular to the orbital plane, zero orbital eccentricity and a simplified geography with the same fraction of continents (0.3) at every latitude, and (iii) a similar pseudo-Earth with a different percentage of continents vs oceans than found on Earth. For each configurations, we compared the latitudinal temperature profiles without vegetation to those with vegetation-albedo feedback, for different values of the semi-major orbital axis and of eccentricity.\\

\indent
The rest of this work is structured as follows. After a schematic summary of the ESTM and vegetation models, the last part of \S \ref{sec:model} illustrates how the two approaches have been analytically combined. Validation of B09's results within ESTM and calibration of the updated Earth's model are summarized in \S \ref{sec:methods}. The following \S \ref{sec:results} is structured into two main parts. After showing how combination with ESTM may implement new pieces of information in the B09 model (\S \ref{subsec:seedprop}), we address the impact of vegetation on the climate and habitability for a terrestrial-type planet as a function of different configurations in the parameter space (\S \ref{subsec:vegalbfeedback}). \S \ref{subsubsec:discussionconclusions} provides a summary and a general discussion of the results, ending up with an overview of possible model extensions and improvements.

\section{The model}
\label{sec:model}

\subsection{Climate Model}
\label{subsec:climmodel} 
The ESTM climate model belongs to the category of EBMs, i.e., it is an Energy Balance Model reproducing the latitudinal dependence of the energy balance between the incoming and outgoing radiation and the energy transport across latitudes. ESTM discretizes the planet surface into $N$ latitudinal bands, and models the meridional heat transport across them by a modified diffusion equation:

\begin{equation}
    C  \frac{\partial T}{\partial t} - 
    \frac{\partial}{\partial x}
    \left[ D \, (1-x^2) \, \frac{\partial T}{\partial x} 
    \right] = [S(1-A(V))] - I. ~
\label{diffusionEq}
\end{equation}

Equation \eqref{diffusionEq} accounts for both the latitudinal heat transport and the heat exchange between the surface, the atmosphere and the outer space. Its different terms and parameters can be understood as follows:

\begin{itemize}
\item
$t$ is orbital time.
\item
$x=\mathrm{sin}(\theta)$ is the sine of the average band latitude.
\item
$T=T(t,x)$ is the latitudinal band temperature.
\item
$C$ represents the effective thermal capacity of the planetary surface per unit area (J m$^{-2}$ K$^{-1}$), calculated by summing up the contributions of lands, oceans, ice over lands, and ice over oceans. Each of these contributions is weighted according to the zonal coverage of each component.
\item
The second term on the left hand describes the meridional energy transport along the coordinate $x$, modulated by the diffusion coefficient $D$ (W m$^{-2}$ K$^{-1}$), which governs the efficiency of the latitudinal heat transport.
\item
On the right hand of Equation \eqref{diffusionEq}, the term $S=S(t,\theta)$ represents the insolation, i.e., the incoming stellar radiation (W m$^{-2}$) which is calculated taking into account the stellar luminosity, the orbital parameters and the inclination of the planet rotation axis.
\item
$A$ is the albedo at the top of the atmosphere (TOA), i.e., the fraction of incoming photons that are reflected back in space without warming up the planet. In the original version of the model (V13, V15), $A$ was specified by the surface distribution of continents, oceans, continental ice, ice on oceans, and clouds. In V13 and V15, a potential (static) contribution of vegetation was also accounted for by lowering the average albedo of land from 0.3$-$0.35 (bare granite; \citep{Dobos2020}) to 0.2 \citep{Williams1997b}. 
\item
$I$ (W m$^{-2}$) is the Outgoing Long-wave Radiation (OLR).
\end{itemize}

Terms $A$ and $I$ are computed using single-column radiative-convective calculations. $A$ is computed as a function of the temperature $T$, the surface albedo, the zenith angle; $I$ is computed as a function of the temperature $T$. Recently, ESTM has been coupled with EOS \citep{Simonetti2022}, allowing the use of both different atmospheric types (with their respective abundances) and different stellar spectra \citep{Biasiotti2022}. A full presentation of the model, including detailed description of the terms $C, D, S, A, I$ that characterize the astrophysical, geophysical and atmospheric properties of the planet can be found in V15.

\subsection{Vegetation Model}
\label{subsec:vegmodel} 
The model for a single vegetation type is based on the introduction of a second partial differential equation, coupled to Equation \eqref{diffusionEq} and describing the growth, death and colonization of vegetation as a function of the temperature $T$. Here, the temperature is taken as a simplified proxy of precipitation as the intensity of the hydrological cycle grows with temperature in the typical range of habitable temperatures, e.g. \citep{vonBloh1997}. The vegetation equation reads:
\begin{equation}
    \frac{\partial V}{\partial t} = 
    s \, \frac{1}{C} \, \frac{\partial}{\partial x}
    \left[D\, (1 - x^{2}) \, \frac{\partial V}{\partial x} \right]
    + G(T)V(1-V) - d(T) V. ~ 
\label{vegetationEq}
\end{equation}

In Equation \eqref{vegetationEq}, $V$ is the fraction of continental surface covered by vegetation, $0\le V \le 1$, $G$ and $d$ are, respectively, the colonization rate of bare areas by vegetation and the vegetation mortality \citep{LentonLovelock2001, Lenton2002, Wilkinson2003, Wood2008, Baudena2009, CrestoAleina2013}, and $s$ represents the seed dispersal term across latitudinal bands. The vegetation growth factor $G(T)$ is described by a truncated parabola (cf. \citep{vonBloh1997, Wood2008}):

\begin{equation}
    G(T) = g_\mathrm{max} \,\, \left\{1 - \left[w \, (T_v - T)\right]^{2} \right\},  ~
\label{logisticEq}
\end{equation}

where  $g_\mathrm{max}$ is the maximum value for the growth factor, $w$ is the width of the parabola, and $T_{v}$ is the optimal temperature for vegetation growth $G(T)$ $-$ which is set to zero for $T<0$. \\

\indent
Following B09, we introduced a hierarchical model (see e.g. \citep{Tilman1994}) describing the competition between trees and grasses, with two stages of tree life: adults and seedlings. The habitat is thus occupied by three types of life forms $-$ adult trees, grasses and tree seedlings. At p. 2, B09 reads: “These three types of life forms are represented by the fraction of space they occupy in the habitat, and thus the model represents spatial dynamics, although implicitly”. The total dynamics is the sum of all these individual
processes: tree seedlings compete for resources with grasses, which can limit their growth by causing local extinction and replacement, while adult trees outcompete grasses. Such model was originally developed in B09 to simulate the dynamics of savanna ecosystems.
In addition, we introduced meridional (tree and grass) seed dispersal.\\

\indent
By assuming that trees do not propagate themselves but generate seeds which, in turn, disperse to produce new seedlings, and grasses generate seeds that also disperse, the vegetation dynamics is described by a set of three differential equations (Eqs. \eqref{baudenaEq1} to \eqref{baudenaEq4}):

\begin{equation}
\frac{\partial V_\mathrm{t}}{\partial t}=c_\mathrm{t}(T)V_\mathrm{ts}-d_\mathrm{t}(T)V_\mathrm{t}; ~
\label{baudenaEq1}
\end{equation}

\begin{equation}
\frac{\partial V_\mathrm{g}}{\partial t}=\left\{ s_\mathrm{g} \, \frac{1}{C}  \,\frac{\partial}{\partial x}\left[D \, (1-x^{2}) \, \frac{\partial V_\mathrm{g}}{\partial x}\right]+c_{\mathrm{g}}(T)V_\mathrm{g}\right\} \cdot
\label{baudenaEq2}
\end{equation}
\[
\hspace{0.7cm}  \cdot \left(1-V_\mathrm{t}-V_\mathrm{g}\right)-d_\mathrm{g}(T)V_\mathrm{g}; ~
\]

\begin{equation}
\frac{\partial V_\mathrm{ts}}{\partial t}=\left\{ s_\mathrm{ts} \, \frac{1}{C} \, \frac{\partial}{\partial x}\left[D \, (1-x^{2}) \, \frac{\partial V_\mathrm{t}}{\partial x}\right]+c_\mathrm{ts}(T)V_\mathrm{t}\right\} \cdot 
\label{baudenaEq3}
\end{equation}
\[
\hspace{0.7cm} \cdot \left(1-V_\mathrm{t}-V_\mathrm{g}-V_\mathrm{ts}\right)-d_\mathrm{ts}(T)V_\mathrm{ts}-c_\mathrm{t}(T)V_\mathrm{ts}-c_\mathrm{g}(T)V_\mathrm{g}V_\mathrm{ts}; ~
\]

\begin{equation}
c_{i}(T)=c_{i}\left\{ 1-\left[w_{i}(T_{i}-T)\right]^{2}\right\}, \hspace{0.5 cm} i=\mathrm{t,g,ts}. ~
\label{baudenaEq4}
\end{equation}

\vspace{0.1cm}
\noindent 
Indices t, g and ts in Eqs. \eqref{baudenaEq1} to \eqref{baudenaEq4} refer, respectively, to trees, grasses and tree seedlings. In this model, $G$ depends on the temperature $T$, and $c$ are the colonization/growth rates as introduced by B09 (in particular, $c_\mathrm{ts}$ measures the colonization success of new areas by tree seedlings generated by adult trees, $c_\mathrm{g}$ is the colonization rate of grasses, and $c_\mathrm{t}$ is the success of tree seedling in growing into adult trees). For a more detailed description of the model and the parameter setup, see B09 and Table \ref{table:vegpar}. For simplicity, we considered vegetation mortality to be independent of temperature. Although $d$  depends on $T$ in the general model of Eq. \eqref{vegetationEq}, after a number of experiments we opted for fixing it for each vegetation specie as indicated in Table \ref{table:vegpar}. 

Using the above model, we allowed competition for resources between two types of vegetation, each having its own albedo and 
growth properties. In general, trees and grasses will have different optimal temperatures $T(i)$, and we shall always choose the widths $1/w(i)$ of the growth rates of the different vegetation types to be large enough to assure the presence of significant competition. Thus, the vegetation distribution and, consequently, the value of the surface albedo is not fixed a priori but it is a result of the competition dynamics controlled by the local temperature which is, in turn, a function also of the albedo. This leads to nonlinear temperature-vegetation feedback which, in turn, can lead to multiple climatic equilibria. Notice, also, that the vegetation considered here only dwells on land. Therefore, the vegetated fraction must be interpreted as a {\it land} fraction; to obtain the total vegetation fraction at a latitudinal band, $V_i$'s has to be multiplied by the continent fraction of the band.

The growth rate dependences adopted here, chosen to  maximize the coexistence between different vegetation types, is different from that made by \cite{Wood2008}. 
In our case, the smaller effect of the difference between the optimal temperatures and the larger widths of the two growth rate curves allow for fully activating the interaction/competition dynamics between the vegetation species (trees and grasses), and lead to a rich dynamical behavior in the parameter space (tree or grass dominance, coexistence, bistability). Indeed, a stronger difference between the growth rate curves would lead to rather obvious conclusions, where at each latitude the outcome of competition would be completely defined by the large difference between the growth rates, and conditions of coexistence or bistability would be extremely limited or absent. With our choice, instead, the competition dynamics can fully reveal itself and can lead to a more complex behavior. Necessarily, the growth rates are limited by a cut-off below zero Celsius (as stated above), and also at larger temperatures. On the other hand, in the planetary configurations to be explored in our work, such larger temperatures are never obtained.

\indent
The vegetation dynamics depends also on seed dispersal. In the model, we considered seed transport as due to the production of the seeds themselves and to their transport by “winds” (modelled in ESTM by the diffusion coefficient $D$). In such sense, the colonization rate at a given latitude  includes the “local” seed propagation inside a latitude band, while the seed dispersion across latitude bands is associated with the long-range meridional atmospheric transport (as represented by $D$) and proportional to the area covered by plants. Clearly, there is the assumption that seed production is proportional to the plant fractional cover.

\indent 
As in V13 and V15, spatial integration of temperature and Eqs. \eqref{diffusionEq} and \eqref{baudenaEq1}$-$\eqref{baudenaEq3} were performed using a staggered grid \footnote{In a few cases, our staggered-grid space integration method may introduce asymmetries in the vegetation profiles. Such cases were thus re-run by doubling the spatial resolution.}.
For the temporal dynamics, we used a fourth order Runge-Kutta integrator. Since we are performing a great number of runs, we adopted an HTC (High-Throughput Computing) approach. To this aim, we wrote a Master/Slave code able to run simultaneously a large number of cases. The code was written in \textsc{Python} (using the library \textsc{mpi4py}), and also implements check pointing and restarting. The runs presented in this work took place on a number of cores ranging from 36 to about 200, on local facilities (e.g., \texttt{amonra@inaf}). Trivially, such a HTC parallelization scales with the number of cores.
A run is considered to have {\it converged} to its equilibrium solution when the global annual average surface temperature does not change by more than $\Delta T /T$ (over two consecutive orbits), and the global annual vegetated fractions of trees, grasses and seedlings do not change by more than $\Delta V / V$. Convergence of the vegetation fractions may be slower $-$ even {\it much} slower $-$ than that of the surface temperature, depending upon the values of vegetation parameters.

\section{Model validation and calibration}
\label{sec:methods}

\subsection{Model validation}
\label{subsec:validation}
\begin{figure*}%
    \centering
   \includegraphics[width=140mm]{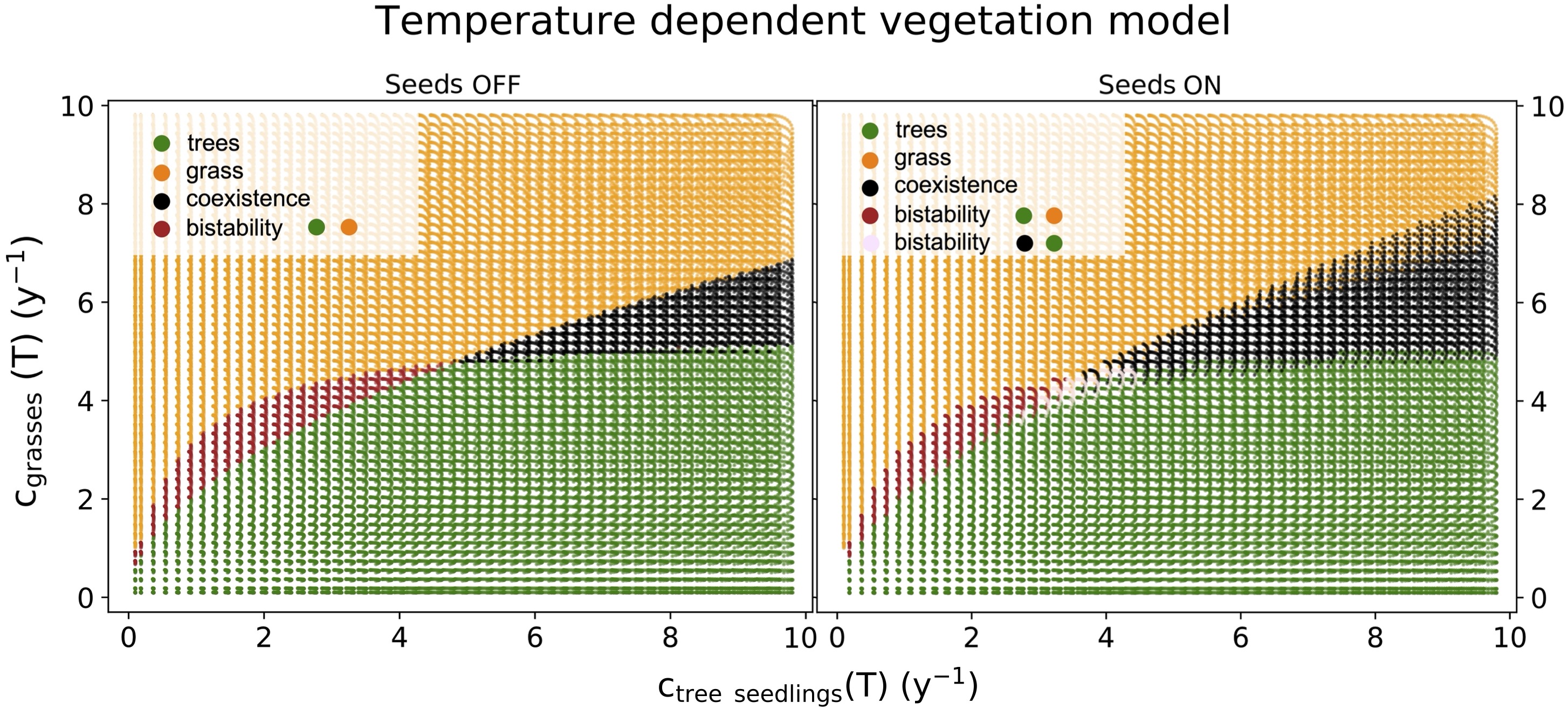}
    \caption{Left panel: Model validation. Reproduction of the results of \citep{Baudena2009} for vegetation growth factors depending on temperature and in the absence of seed transport. Right panel: Extension of the Baudena's model by the introduction of seed transport.}
    \label{fig:BauMaps}
\end{figure*}

The model was validated by reproducing the equilibrium states in the subset of  parameter space spanned by the colonization rate of grasses and tree seedlings, $c_\mathrm{g}$ and $c_\mathrm{ts}$ respectively, while $c_\mathrm{t}$ was kept fixed to 0.2 $-$ according to B09 (Fig. 1 therein and Fig. \ref{fig:BauMaps}, left panel, herein). At the stage of model validation, vegetation-albedo feedback was turned off and seed transport was set to zero (that is, $s_\mathrm{t}=s_\mathrm{g}=s_\mathrm{ts}=0$). The first choice implied that the temperature profile at the equilibrium only depended on planetary parameters, i.e., not on vegetation $-$ which played a passive role in all the validation runs. This was obtained by setting  $A(V)=A$ in Equation \ref{diffusionEq}. Here $A$ on land is kept fixed as in V13, and does not depend on the vegetation fraction.

In our model, colonization rates depend on temperature through Equation \ref{baudenaEq4}, and for each temperature provided by ESTM in a given latitudinal band, we have a unique set of colonization rates once the functional dependencies of the vegetation parameters are chosen. Thus, a given latitudinal band in a given ESTM run provides a set of colonization rates that define a single point in the parameter plane of Fig. \ref{fig:BauMaps}. By producing several runs of the ESTM for  different choices of the optimal temperatures (for passive vegetation), and checking what type of vegetation behavior was obtained when stationarity was reached, we populated the parameter space of colonization rates with the corresponding asymptotic vegetation behavior. All other parameters were set as reported in Tables \ref{table:ESTMpar} and \ref{table:vegpar}, respectively for the climate and the vegetation model.
In particular, each simulation started from initially constant values of $T$ = 275 K, $c_\mathrm{g}$ = 0.9 and $c_\mathrm{ts}$ = 0.09. As for bistability, the last two values were inverted (values in brackets at the top  of Table \ref{table:vegpar}).
The choice of adopting a totally symmetrical geographic configuration, with orbit eccentricity and axis inclination both set to zero (the planet setup herein is named \textit{pseudo-Earth}) is motivated by the requirement of having a simple, symmetric surface temperature profile without seasonal variations for the purpose of model validation.\\

\begin{table}%
\centering
\caption{Model validation: ESTM parameters for our {\it pseudo-Earth} configuration.}
\label{table:ESTMpar}
\scalebox{0.8}{%
\footnotesize
\begin{tabular}{l|ll}%
\hline\hline
\textbf{Parameter} & \textbf{Value} & \textbf{Comment}\\
\hline
    smaP & 1.0 AU & Semi-major axis of the planet\\
	eccP & 0.0 & Eccentricity of the planet's orbit\\
	obliq & 0.0 deg & Obliquity of the planet rotation axis\\
	f$_\mathrm{0}$$_\mathrm{const}$ & 0.70 & Constant fraction of ocean\textsuperscript{1}\\
	a$_{\text{sl}}$ & 0.20 & Surface albedo of lands\textsuperscript{1}\\
	a$_{\text{sil}}$ & 0.70 & Surface albedo of ice on lands\\
	a$_{\text{sio}}$ & 0.70 & Surface albedo of ice on ocean\\
	f$_{\text{cw}}$ & 0.70 & Cloud coverage on water\textsuperscript{2}\\
	f$_{\text{cl}}$ & 0.60 & Cloud coverage on lands\textsuperscript{2}\\
	f$_{\text{ci}}$ & 0.60 & Cloud coverage on ice\\
\hline\hline
\multicolumn{3}{l}{%
  \begin{minipage}{6.5cm}%
    \tiny \footnotemark[1]{As in V13.} \hspace{0.1cm} \footnotemark[2]{As in V15.}%
  \end{minipage}%
}\\
\end{tabular}}
\end{table}

\begin{table}%
\centering
\caption{Model validation: ESTM parameters for vegetation. Seed dispersal is set to zero for model validation, to 0.1 afterwards.}
\label{table:vegpar}
\scalebox{0.8}{%
\footnotesize
\begin{tabular}{l|ll}%
\hline\hline
\textbf{Parameter} & \textbf{Value} & \textbf{Comment}\\
\hline
    V$_{\text{t\_start}}$ & 0.01 (0.9) & Initial fraction of trees\\
	V$_{\text{g\_start}}$  & 0.9 (0.01) & Initial fraction of grasses\\
	V$_{\text{ts\_start}}$  & 0.09 & Initial fraction of tree seedlings\\
	T(V$_{\text{t}}$) & 285 K & Optimal temperature for trees\\
	T(V$_{\text{g}}$) & 304 K & 
    Optimal temperature for grasses\\
	T(V$_{\text{ts}}$) & 285 K & 
    Optimal temperature for tree seedlings\\
	c$_{\text{t}}$ & 0.2 & Growth factor for trees\textsuperscript{1}\\
	d$_{\text{t}}$ & 0.02 & Mortality for trees\textsuperscript{1}\\
	d$_{\text{g}}$ & 0.5 & Mortality for grasses\textsuperscript{1}\\
	d$_{\text{ts}}$ & 0.5 & Mortality for tree seedlings\textsuperscript{1}\\
	w$_{\text{t}}$ & 0.01 & 
    Logistic width for trees\\
	w$_{\text{g}}$ & 0.01 & 
    Logistic width for grasses\\	
    w$_{\text{ts}}$ & 0.01 & 
    Logistic width for tree seedlings\\
	s$_{\text{g}}$ & 0.0 (0.1) & Seed dispersal coefficient for grasses\\
	s$_{\text{ts}}$ & 0.0 (0.1) & Seed dispersal coefficient for tree seedlings\\
\hline\hline
\multicolumn{3}{l}{%
  \begin{minipage}{6.5cm}%
    \tiny \footnotemark[1]{As in B09.}%
  \end{minipage}%
}\\
\end{tabular}}
\end{table}

The left panel of Fig. \ref{fig:BauMaps} shows the results of validation (for the case without seed transport; Seeds OFF). Each run provided a triplet of vegetated land fractions (trees, grass and tree seedling), one for each latitudinal band. Dependence of trees/grasses vegetation fractions on latitude are exemplified in Fig. \ref{fig:LatProfiles}. As a consequence, each run produced many points ($\leq 54$) in Fig. \ref{fig:BauMaps}, one per each band where vegetation fractions were not null. 
As in B09, we found that the system can display four different types of behavior: in some portions of parameter space only trees are found (forest, green points in the figure), while  other parameter values led to a dominance of grasses (grassland, yellow points). On the other hand, stable coexistence of grasses and trees (black points) is also possible. Here, we requested a value of at least 1 per cent for the fraction of area occupied by each vegetation type for declaring coexistence. For other parameter sets, bistability (red points) between grassland and forest was observed. In bistable cases, the system converged either to a completely herbaceous state (grassland) or to a woody equilibrium (forest) depending on the initial value of the tree and grass fractions (set in Table \ref{table:vegpar}). Higher grass colonization rates make the planet grassland-dominated; conversely, when tree seedlings enhance their ability to occupy new space, a forested planet is formed. For intermediate values of the colonization parameters, either coexistence or bistability were found.
These results fully agree with those of B09 \footnote{In the absence of feedback, on the \textit{pseudo-Earth} the temperature converges to its equilibrium latitudinal profile after 25 orbits. We here adopted a `cold start', i.e., we set the initial temperature at a constant value of 275 K over the entire planet. However, as the system of differential equations is such that at $T < 0^{\circ}$C vegetation is prevented from growing anymore and carries over death (as specified above), it may happen that during this transient it completely disappears and, in the absence of seed transport, cannot be restored $-$ especially when the initial vegetated fraction is low. Surface temperatures can therefore temporarily drop below 0$^{\circ}$C, especially at high latitudes. This effect can make it appear some anomalous bistable points: not only transitions from trees to grasses, and/or vice versa, but also from coexistence between trees and grasses to trees, or grasses. To correct such anomalies, the adopted solution was freezing vegetation evolution during the first 25 orbits.}.

\subsection{Seed dispersal}
\label{subsec:seedprop}
The right panel of Fig. \ref{fig:BauMaps} shows how the parameter space reproduced in the validation stage is affected by seed dispersal (Seeds ON).
The parametric setup is the same of Table \ref{table:ESTMpar} and Table \ref{table:vegpar}, apart from the seed dispersal coefficients that were set to 0.1 (values in brackets at the bottom of Table \ref{table:vegpar}).
The effects of seed dispersal are: (i) widening the region of coexistence between trees and grasses; (ii) changing the shape of the region of bistability; (iii) introducing a new type of bistability. As for (i), trees can now propagate to regions that were previously occupied only by grasses. This is due to the fact that, at high latitudes (expressed as absolute values), the new source term in Equation \ref{baudenaEq3}:

\begin{equation}
s_\mathrm{ts} \, \frac{1}{C} \, \frac{\partial}{\partial x}\left[D \, (1-x^{2}) \, \frac{\partial V_\mathrm{t}}{\partial x}\right] ~
\label{baudenaEq1b}
\end{equation}

\vspace{0.1cm}
\noindent
allows trees to survive in cases where a tree-dominated initial condition is used. At high latitudes, this transforms the equilibria at points ($c_\mathrm{g}$, $c_\mathrm{ts}$) (corresponding to grass dominance without seed propagation) into equilibria corresponding to trees. Although less significantly, an analogous effect is observed for grasses at the lower border of the coexistence region. The effect (ii) can be explained similarly. (iii) A new type of bistability appears around $c_\mathrm{g} \approx 4$ and $c_\mathrm{ts} \approx 4$ (despite $^2$). Here, for an initial condition of grass dominance, the system evolves towards coexistence $-$ thanks to the same mechanism which extends the coexistence region below its lower border. 
This leads to bistability between coexistence and tree dominance $-$ a new behaviour with respect to the original formulation of the B09 model without seed dispersal.\\

\begin{table}
\centering
\caption{Model calibration for Earth conditions: ESTM parameters\textsuperscript{1,2} and vegetation albedo values.}
\label{table:vegcal}
\scalebox{0.8}{%
\footnotesize
\begin{tabular}{l|ll}%
\hline\hline
\textbf{Parameter} & \textbf{Value} & \textbf{Comment}\\
\hline
	eccP & 0.01671022 & Eccentricity of the planet's orbit\\
	obliq & 23.43929 deg & Obliquity of the planet rotation axis\\
	a$_{\text{sl}}$ & 0.35 & Surface albedo of lands\textsuperscript{2}\\
	s$_{\text{t}}$ & 0.1 &  Seed dispersal coefficient for grasses\\
	s$_{\text{ts}}$ & 0.1 & Seed dispersal coef. for tree seedlings\\
	a$_{\text{t}}$ & 0.073 & Surface albedo of trees\\
	a$_{\text{g}}$ & 0.24 & Surface albedo of grasses\\
	a$_{\text{ts}}$ & 0.073 & Surface albedo of tree seedlings\\
\hline\hline
\multicolumn{3}{l}{%
  \begin{minipage}{6.5cm}%
    \tiny \footnotemark[1]{As in V13.} \hspace{0.1 cm} \footnotemark[2]{As in V15.}%
  \end{minipage}%
}\\
\end{tabular}}
\end{table}

\begin{figure*}
    \centering
    \includegraphics[width=78mm]{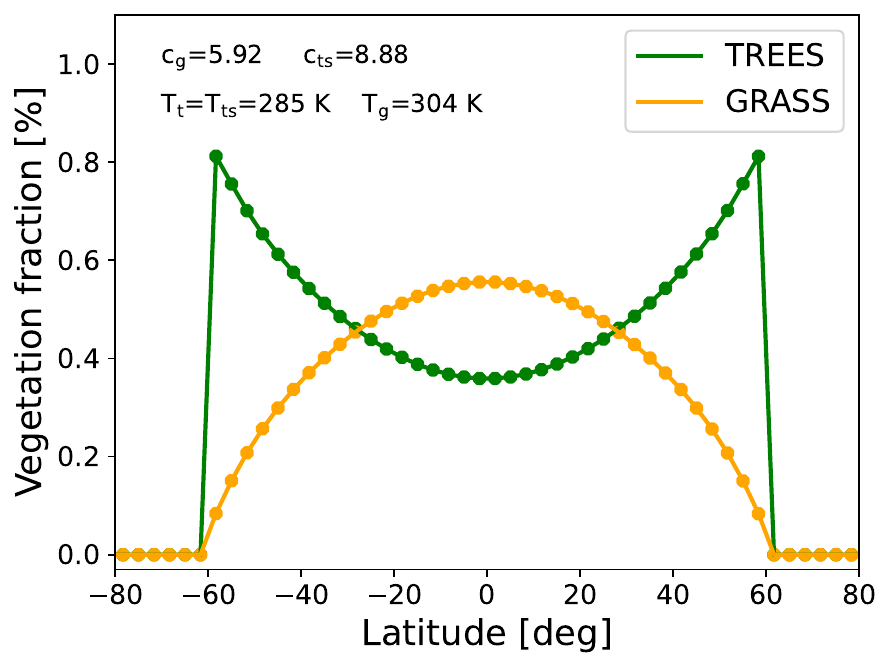}
    \hspace{0.5cm}
    \includegraphics[width=78mm]{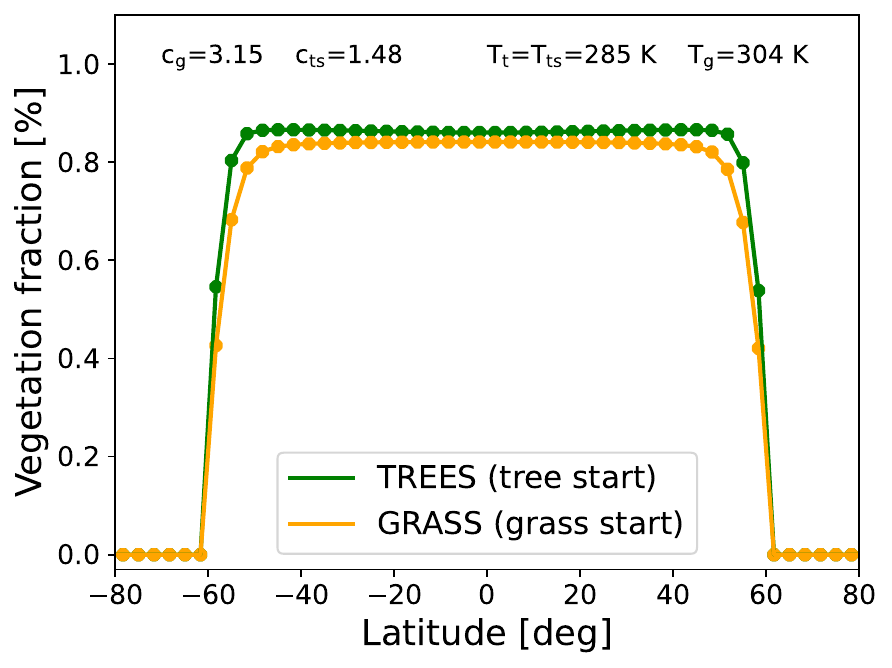}
    \includegraphics[width=75mm]{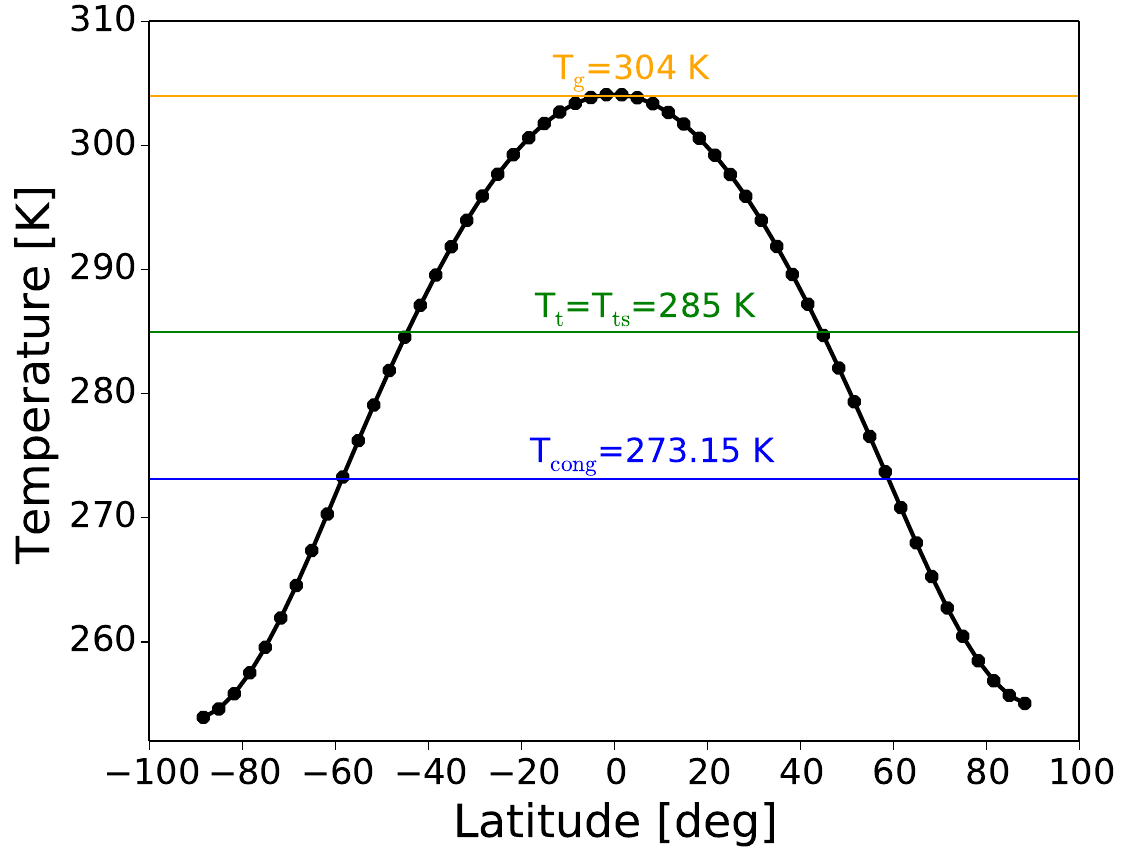}
    \caption{Upper panels: examples of latitudinal vegetation profiles for trees (green/dark lines) and grasses (yellow/light lines): left panel corresponds to a case of coexistence; right panel corresponds to a case of bistability. Optimal temperatures for vegetation are as in Fig. \ref{fig:BauMaps}. Lower panel: temperature profile. All graphs refer to the \textit{pseudo-Earth} configuration.}
    \label{fig:LatProfiles}
\end{figure*}

Upper panels of Fig. \ref{fig:LatProfiles} show the latitudinal profiles for two representative cases of vegetation fraction $-$ respectively corresponding to a coexistence and a bistable case, for the \textit{pseudo-Earth} configuration.
For coexistence with initial conditions of grass dominance (upper left panel), trees prevail at high latitudes (boreal forests), while grasses prevail at low  latitudes. According to our choice, the optimal temperatures of trees and grasses differ by 20 degrees (see Table \ref{table:vegpar}). Such a choice, combined with the relatively wide logistic function for the temperature dependence of colonization rates, extends the temperature range (and thus the latitudinal range) where trees and grasses can coexist in the same area. A different choice, i.e., a larger optimal temperature difference and/or a narrower logistic function would have produced only trees at high latitudes and only grasses near the Equator. Movies with the complete set of profiles for all possible configurations in the ($c_\mathrm{g}$, $c_\mathrm{ts}$) parameter space are accessible at the links provided in Appendix A.\\ 
\indent
The lower panel of Fig. \ref{fig:LatProfiles} provides the temperature profile for the same \textit{pseudo-Earth} configuration; being this a case characterized by null obliquity and eccentricity, such temperature profile corresponds to the annual average. Green and yellow lines respectively illustrate the optimal temperature for trees and grasses. Since in these runs vegetation is “passive”, vegetation albedo does not influence the surface temperature; such fractions only depend on $c(t)$ of trees and grasses.

\subsection{Model calibration}
\label{subsec:calibration}
\noindent
The full climate-vegetation model was calibrated for Earth conditions, by selecting a point of coexistence in the validation map (i.e., $c_\mathrm{g}$ = 5.92 and $c_\mathrm{ts}$ = 8.88 in the left panel of Fig. \ref{fig:BauMaps}, corresponding to a concentration of trees plus tree seedlings approximately equal to that of grasses). Our target was to find out a triplet of albedo values able to reproduce, within the full model (i.e., after having switched on both seed transport and vegetation feedback), the correct global annual average temperature. We then looked for the values of vegetation albedo (respectively, for trees, grasses and tree seedlings) able to reproduce the global annual average temperature as calculated by the previous version of ESTM (that is, 289 K; V15), where the land surface albedo was decreased from that of pure granite (0.35; \cite{Dobos2020}) to a lower value accounting for a contribution by generic vegetation (0.2; V13). Here, on non-vegetated (bare) land we adopted the albedo value of granite, thus the overall land albedo resulted from the combination of vegetation and bare land albedo. Given the non-unique combinations of trees and grass albedo values able to achieve such a result, we explored different setups. Our choice here, illustrated in Table \ref{table:vegcal}, assumes grass to be lighter and trees darker, in keeping with standard terrestrial values (e.g., \href{https://www.climate.be/textbook/chapter1_node16.html}{https://www.climate.be/textbook/chapter1$\_$node16.html}). When not differently specified, planetary parameters in Table \ref{table:vegcal} are the same as in Table \ref{table:ESTMpar}.

\section{Results}
\label{sec:results}

\begin{table}
\centering
\caption{Parametric setup for the three planetary configurations explored in this study: \textit{Earth}, \textit{pseudo-Earth}, \textit{dry pseudo-Earth}.}
\label{table:threefeedbackpar}
\scalebox{0.8}{%
\footnotesize
\begin{tabular}{l|ccc}%
\hline\hline
\textbf{} & \textbf{Earth} & \textbf{pseudo-Earth} & \textbf{dry pseudo-Earth}\\
\hline
	eccP & 0.01671022 & 0.0 & 0.0\\
	obliq & 23.43929 deg & 0.0 deg & 0.0 deg\\
	geography & present Earth\textsuperscript{1} & modified\textsuperscript{2} & modified\textsuperscript{2}\\
	f$_\mathrm{0}$$_\mathrm{const}$ & 0.70 & 0.70 & 0.30\\
	s$_{\text{t}}$ & 0.0 & 0.0 & 0.0\\
	s$_{\text{g}}$ & 0.1 & 0.1 & 0.1\\
	s$_{\text{ts}}$ & 0.1 & 0.1 & 0.1\\
\hline\hline
\multicolumn{4}{l}{%
  \begin{minipage}{7.5cm}%
    \tiny \footnotemark[1]{Sampled over 46 latitude strips and interpolated to the desired number of strips; \citep{Murante2020}.}\\
    \tiny \footnotemark[2]{Constant fraction of oceans at all latitude bands; \citep{Murante2020}.}%
  \end{minipage}%
}\\
\end{tabular}}
\end{table}

\subsection{Vegetation-albedo feedback}
\label{subsec:vegalbfeedback}
The main goal of this study was to understand whether and how the vegetation-albedo feedback influences planetary habitability. As anticipated above, here this effect was simulated through the Charney mechanism: as vegetation is darker than the bare surface of the continents, it impacts the surface albedo and therefore warms up the planet. In our model, turning on vegetation-albedo feedback means using the full Eqs. \eqref{baudenaEq1} to \eqref{baudenaEq4}, thus allowing the albedo on land to depend on vegetation fractions. 
Three planetary configurations were considered: (i) the \textit{Earth}, as modeled by V15; (ii) a \textit{pseudo-Earth} with its rotation axis perpendicular to the orbital plane, zero orbital eccentricity and a simplified geography (constant fraction of oceans at all latitudinal bands, $f_\mathrm{0}=0.7$); and (iii) a \textit{dry pseudo-Earth} differing from the pseudo-Earth in the fraction of continents vs oceans ($f_\mathrm{0}=0.3$). The Earth-like configuration is motivated by the need for testing, as a benchmark, the impact of the simplified vegetation model adopted here. The other two configurations (pseudo-Earth and dry pseudo-Earth) have the advantage of simplifying interpretation of results, by removing complications such as, for instance, seasonal effects and/or planetary geography. In particular, the dry pseudo-Earth configuration amplifies the effect of the Charney mechanism, increasing the faction of lands with respect to oceans.
Vegetation growth below the water freezing point was prevented by setting to zero the seed dispersion term whenever the temperature of a latitudinal band fell below $T = 0$$^{\circ}$C. 
Initial tree, grass and tree seedlings' fractions over land were respectively initialized at 9, 90 and 1 per cent, consistently with previous calculations. Different parametric setups for these three configurations, when different from those reported in Table \ref{table:ESTMpar} and Table \ref{table:vegpar}, are summarized in Table \ref{table:threefeedbackpar}.

\subsubsection{The Earth as a benchmark}
\label{subsubsec:earth}

\begin{figure*}
    \centering
    \includegraphics[width=136.5mm]{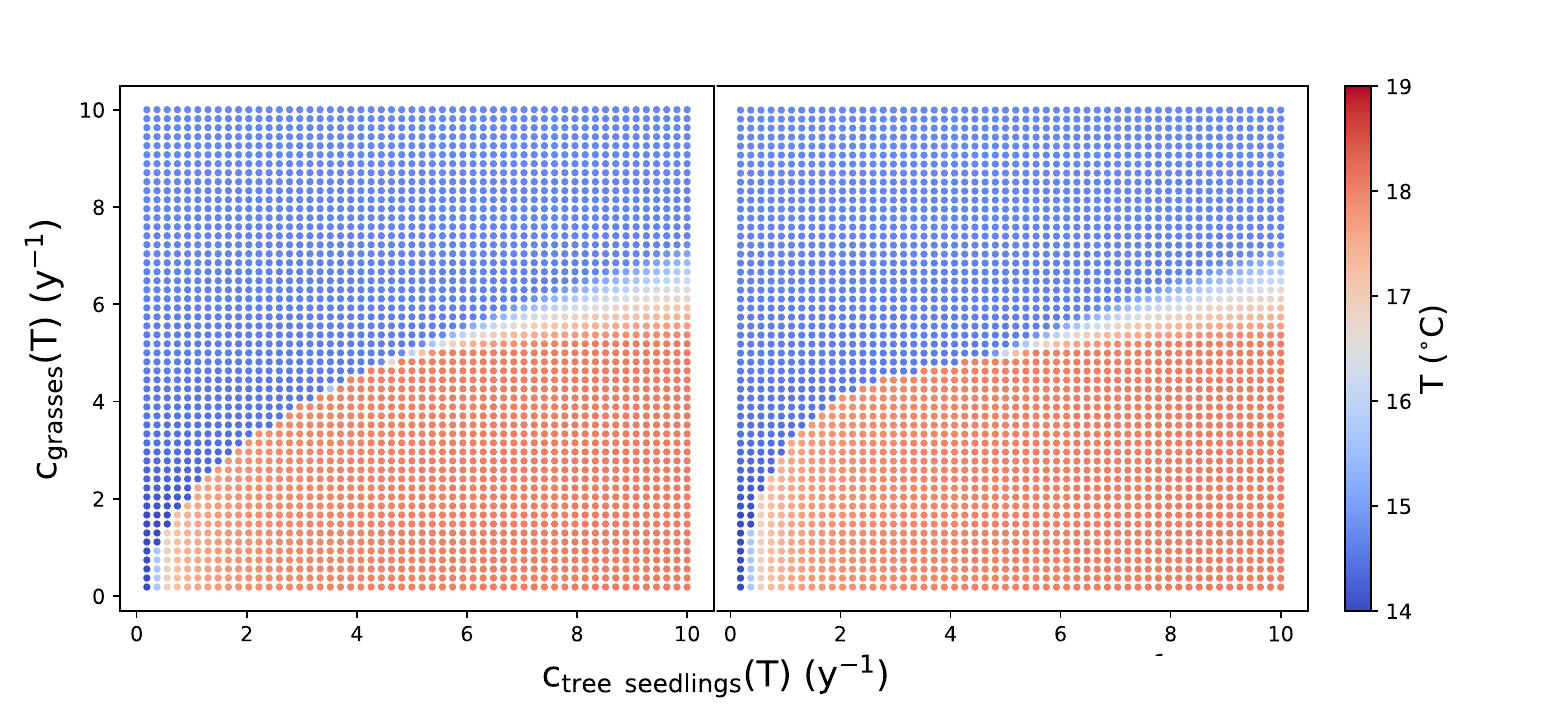}
    \hspace{0.5cm}
    \includegraphics[width=84mm]{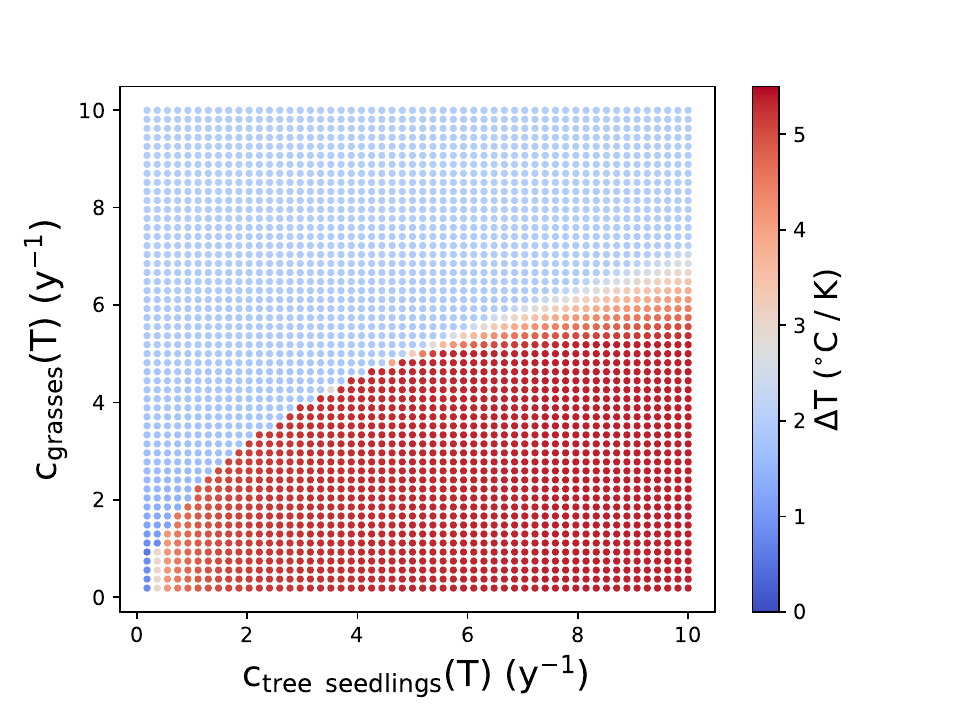}
    \includegraphics[width=84mm]{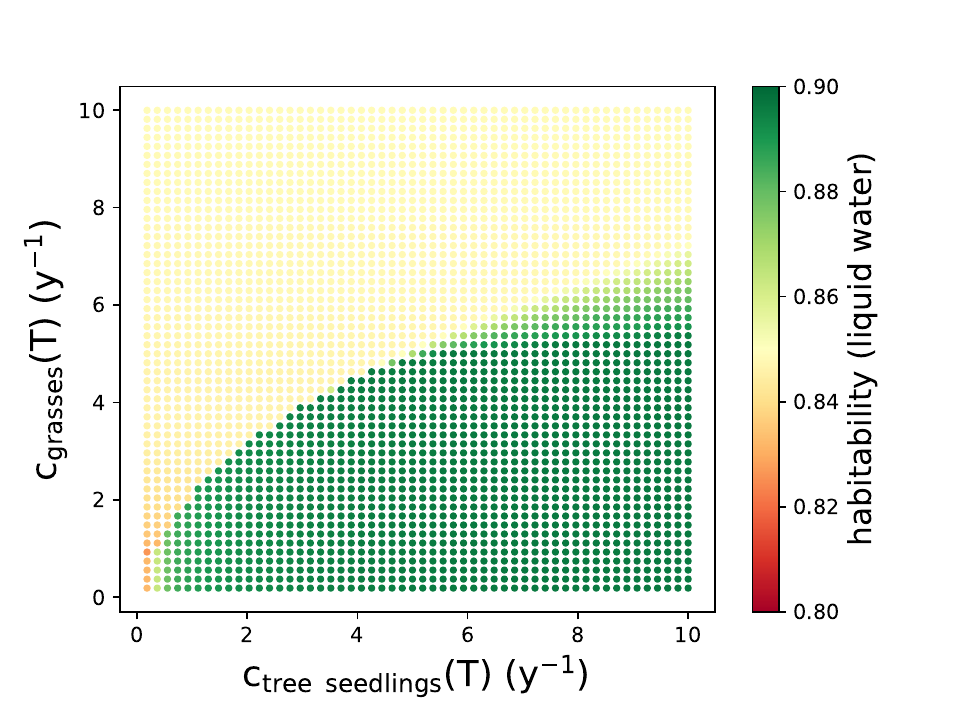}
    \caption{Global annual average temperature following different (and symmetrical) initial conditions (upper panels), temperature anomaly (bottom left panel), and liquid-water habitability (bottom right panel) for the {\it Earth}, as a function of different combinations of growth factors for tree seedlings and grasses. For comparison, the habitability of Earth with bare granite continents is 0.818. Color scales show: the global annual average temperature (in Celsius); the anomaly, defined as the difference between the global annual average temperature of each specific run and that of a planet with bare granite continents and no vegetation cover (in Celsius or, equivalently, in Kelvin); the liquid-water habitability, defined as a pure number comprised between 0 and 1 (according to V15).}
    \label{fig:Earth}
\end{figure*}

The planetary parameters for the Earth configuration are listed in the second column of Table \ref{table:threefeedbackpar}, where calibration of the vegetation albedo was performed for $c_\mathrm{g} = 5.92$ and $c_\mathrm{ts} = 8.88$.
Upper panels of Fig. \ref{fig:Earth} show the global annual average temperature for each of the 54 $\times$ 54 numerical runs, after vegetation-albedo feedback had been turned on. As in Fig. \ref{fig:BauMaps}, each run was initialized by setting each $c_\mathrm{g}$ and $c_\mathrm{ts}$ to the values corresponding to their optimal temperatures. For the sake of completeness,  the effect of bistability on the surface temperature is also shown. When initial conditions return a grass-dominated solution, the temperature is lower; conversely, when $-$ for the same values of the growth factor $-$ initial conditions return a tree-dominated configuration, temperature there is higher. Bottom left panel of Fig. \ref{fig:Earth} shows the temperature anomaly with respect to the Earth without vegetation on land, i.e., the difference between the global annual average temperature of each run and that of Earth having the albedo of granite on the continents. The results indicate that both trees and grasslands can significantly impact the global annual average temperature. On average, the Earth warms by about 1.5$-$2 degrees with respect to the granite case for grasslands, and up to 5 degrees for tree dominance. Since grasslands feature the highest albedo, they have a smaller impact; a larger effect occurs for tree-covered continents. Intermediate behaviors are observed in the tree-grass coexistence region. 
At the bottom right of Fig \ref{fig:Earth}, we show planetary liquid-water habitability, $h$, defined here as the fraction of planetary surface (averaged over an entire orbit) whose temperature is between the water freezing and evaporation points (V15). For comparison, the habitability of Earth with granite-covered continents is $h=0.818$ and that of the calibration case without vegetation feedback (constant land albedo intermediate between vegetation and granite) is $h=0.877$.
The fact that the overall habitability  maps closely follow those of the global average temperature is a direct consequence of how habitability is defined. Regarding the dependence on the initial conditions (bistability), observed behaviors concerning the latter plots are analogous to what is shown in the upper panels of Fig. \ref{fig:Earth} and here omitted in order to avoid redundancy. The same applies to the next two figures.

\subsubsection{Pseudo-Earth}
\label{subsubsec:pseudoearth}

\begin{figure}
    \centering
    \includegraphics[width=84mm]{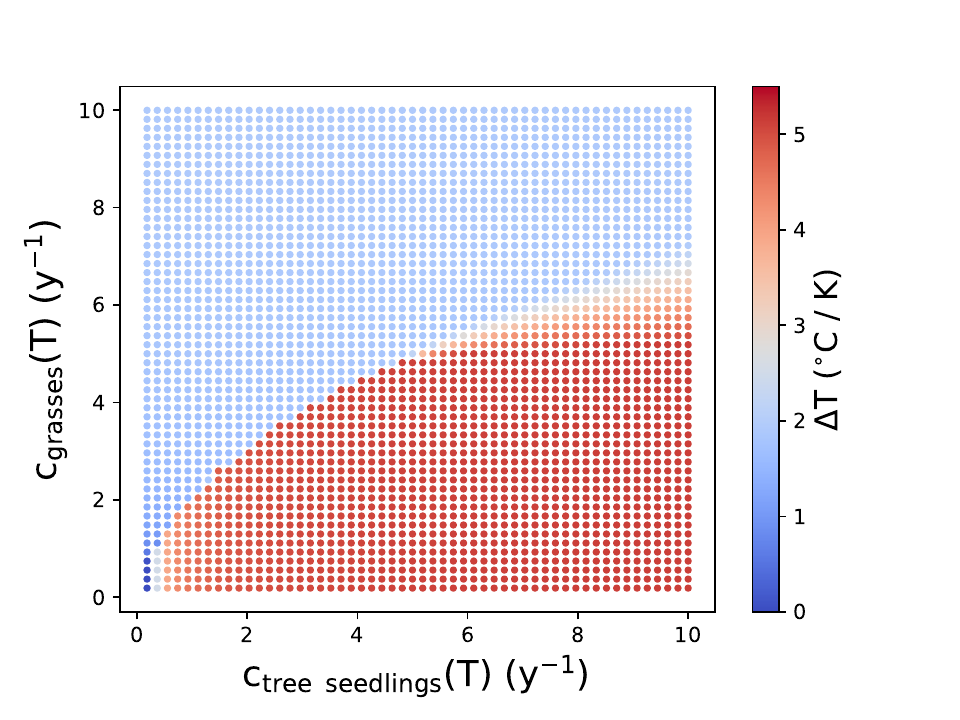}
    \includegraphics[width=84mm]{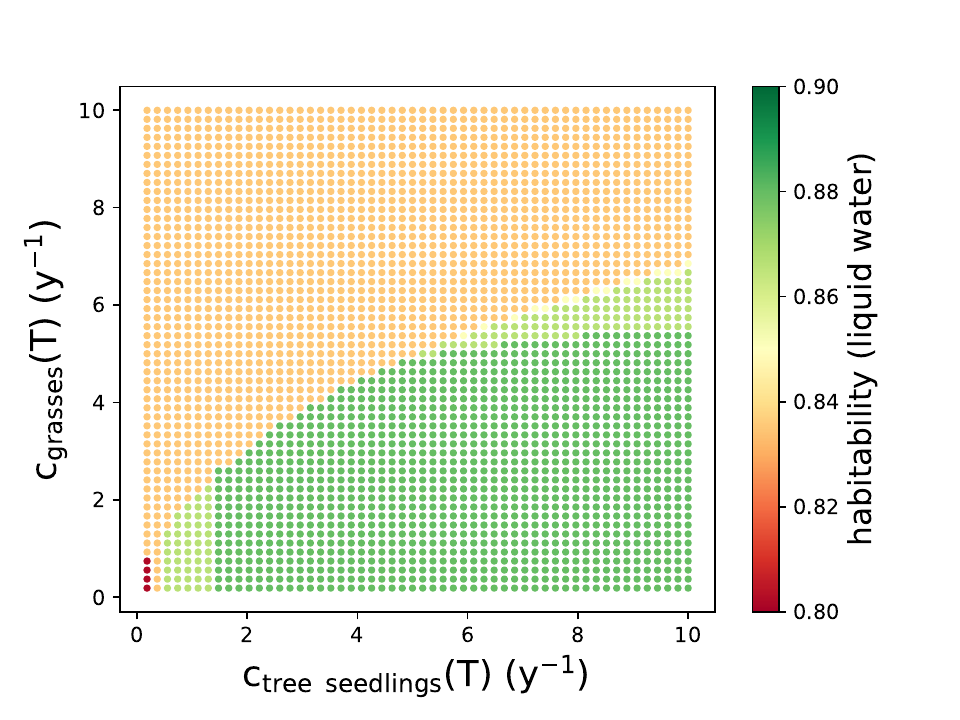}
    \caption{Global annual temperature anomaly (upper panel) and liquid-water habitability (lower panel) for the {\it pseudo-Earth}, as a function of different combinations of growth factors for tree seedlings and grasses. For comparison, the habitability of pseudo-Earth with bare granite continents is 0.802. Color scales as in Fig. \ref{fig:Earth}.}
    \label{fig:PseudoEarth}
\end{figure}

The third column of Table \ref{table:threefeedbackpar} reports the planetary parameters for the pseudo-Earth, where all simulations have been initialized as for the Earth.
The results of the simulations are shown in Fig. \ref{fig:PseudoEarth}. Here, the upper panel shows the temperature anomaly between the case with vegetation feedback and that with constant granite albedo on continents, while the lower panel show the liquid-water habitability.\\
\indent
For comparison, the habitability of pseudo-Earth without vegetation cover is 0.802. The pseudo-Earth behaves similarly to the Earth. Since planetary parameters are different, the annual global average temperature also does, with a tendency of being higher than on Earth $-$ an effect mainly related to the more homogeneous continental cover. As a consequence, the coexistence region becomes warmer. As for the Earth, grass-dominated worlds are colder than tree-dominated planets. The deviation with respect to the bare-rock run, as well as the impact of vegetation on habitability, are similar to what happens on Earth. In general, however, pseudo-Earth habitability is slightly lower. This depends on Earth's seasonal variability, which implies that high latitude bands may be frozen for only a fraction of the year. Differently, on the pseudo-Earth such bands are always frozen $-$ making the overall habitability lower. Note that the discrete blocks visible in the lower panel of Fig. \ref{fig:PseudoEarth} are generated by the dichotomy of pseudo-Earth's latitudinal bands, which are either always habitable or always non-habitable. 
Thus, this lower habitability is affected more by the lack of seasons than by the higher homogeneity in the land distribution: indeed, the latter pushes average temperature, and therefore habitability, towards higher values.

\subsubsection{Dry pseudo-Earth}
\label{subsubsec:pseudoearthdry}

\begin{figure}
    \centering
    \includegraphics[width=84mm]{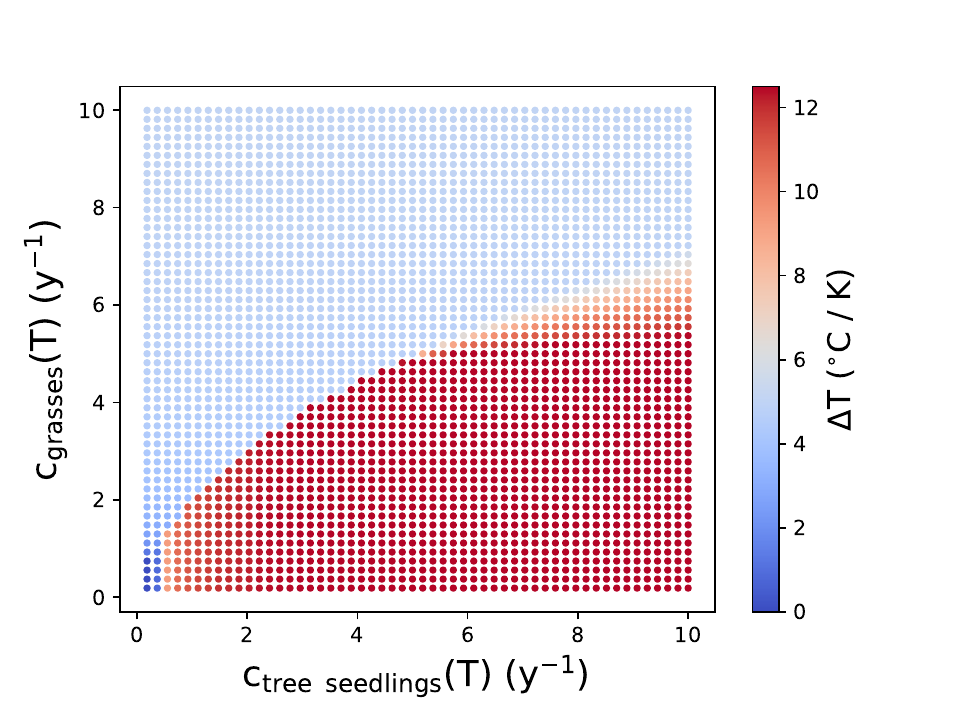}
    \includegraphics[width=84mm]{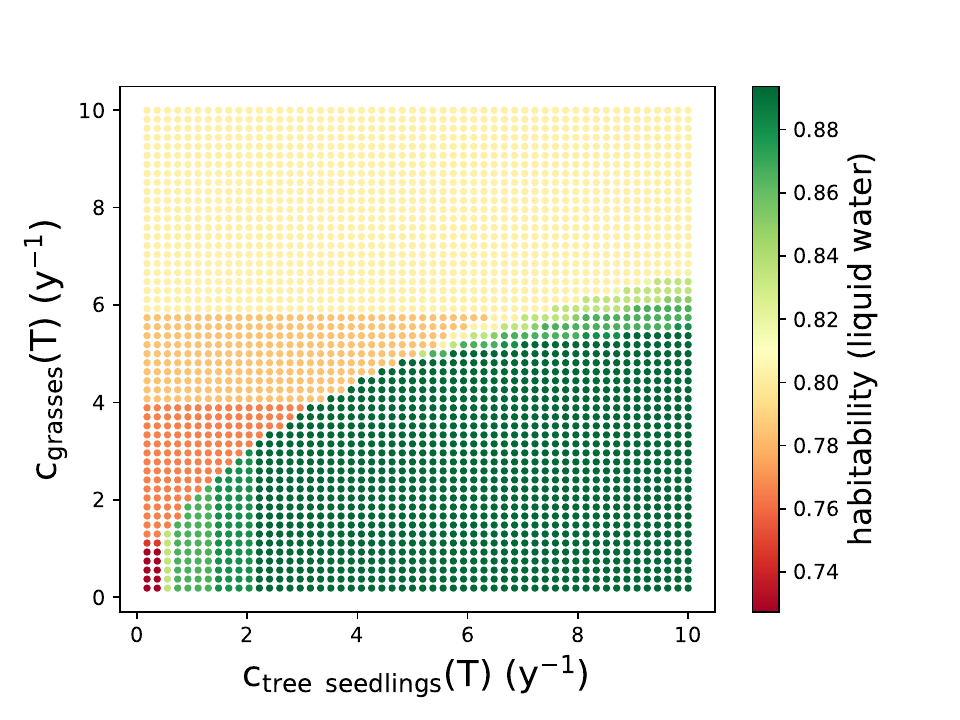}
    \caption{Global annual temperature anomaly (upper panel) and liquid-water habitability (lower panel) for the {\it dry pseudo-Earth}, as a function of different combinations of growth factors for tree seedlings and grasses. For comparison, the habitability of dry pseudo-Earth with bare granite continents is 0.727. Color scales are as in Fig. \ref{fig:Earth}, although their ranges differ due the large variability introduced by the present setup.}
    \label{fig:DryPseudoEarth}
\end{figure}

Fig. \ref{fig:DryPseudoEarth} refers to the dry Pseudo-Earth configuration, whose parameters are listed in the fourth column of Table \ref{table:threefeedbackpar}.
Here, the only difference with respect to the pseudo-Earth is that oceans occupy only 30 per cent of the planetary surface. 
Since the planetary surface occupied by vegetation is larger, its impact on the average global temperature is visibly reinforced, with an increase depending on the type and amount of vegetation cover. Tree-dominated worlds are up to 12 degrees hotter than the bare-rock case with no vegetation; grass-dominated worlds are about 5 degrees warmer. As for the Earth and the pseudo-Earth, the coexistence region corresponds to intermediate values between these two values. 
This makes the impact of vegetation on the climate of dry worlds somehow extreme as far as the Charney mechanism is concerned, as should be expected by the original motivation of Charney's work on desert ecosystems. Again, surface temperature is mirrored by habitability, which now spans a wider range of values (here, the habitability of dry pseudo-Earth with bare granite continents is 0.727). Note how, in general, the habitability of these dry worlds is {\it lower} than that of pseudo-Earths. This is because their larger land cover triggers an increase in the albedo and therefore cools down the planet, even in the presence of trees, as the albedo of oceans is generally lower than that of lands. As for the pseudo-Earth, lack of seasons is responsible for the discrete blocks between separate regions in the parameter space (lower panel of Fig. \ref{fig:DryPseudoEarth}).

\subsubsection{Impact of vegetation on the Circumstellar Habitable Zone}
\label{subsec:CHZ}
Figs. \ref{fig:OuterEdgeEarth}, \ref{fig:OuterEdgePseudoEarth} and \ref{fig:OuterEdgePseudoDryEarth} show, respectively for the three planetary setups reported in Table \ref{table:threefeedbackpar}, the impact of vegetation on the liquid-water habitability index $h$ (as defined in \S \ref{subsubsec:earth}) as a function of the semi-major axis $d$ (for these configurations, equivalent to the distance from the star) and for the following conditions on the vegetation growth factors: dominance of trees ($c_\mathrm{g}(T)$ = 2.035; $c_\mathrm{ts}(T)$ = 9.435); dominance of grasses ($c_\mathrm{g}(T)$ = 8.14; $c_\mathrm{ts}(T)$ = 1.295); coexistence between trees and grasses ($c_\mathrm{g}(T)$ = 5.92; $c_\mathrm{ts}(T)$ = 8.88). 
For Earth-like conditions, the semi-major axis $d$ = 1.04 UA represents the turning point where vegetation enhances habitability from $h$ = 0.0 to $h$ = 0.485 (in the grass-dominance case), $h$ = 0.584 (for coexistence), and $h$ = 0.612 (in the tree-dominance case). This corresponds to the transition from a fully-developed snowball to a planet with intermediate habitability. A similar effect is observed at $d$ = 1.052 for the pseudo-Earth, and at $d$ = 1.035 for the dry pseudo-Earth. The latter case is also the one featuring the highest discrepancy between the profiles associated to different biome compositions (Fig. \ref{fig:OuterEdgePseudoDryEarth}), again as a consequence of the highest fraction of continents.\\

\indent
This effect is further explored in Fig. \ref{fig:OuterEdgePseudoDryEarthLat}. Here, we compare the temperature profiles without and with vegetation as a function of the planetary latitude, for the case of \textit{dry pseudo-Earth}. Again, tree-dominance, grass-dominance, and coexistence are considered. For the same values of the semi-mayor axis, when vegetation is absent, the planet is in a snowball state; by contrast, vegetation generates an equatorial habitable belt, where liquid water can persist. This suggests that a living planet, characterized by a well-developed biosphere, can effectively prevent the onset of a full snowball state.

\section{Discussion and Conclusion}
\label{subsubsec:discussionconclusions}
In this work, we explored the role of vegetation albedo in determining the surface temperature of a rocky planet and, as a consequence, planetary habitability.  
By considering different combinations of growth factors for trees and grasses, four situations emerged, consistent with B09: complete tree dominance (forest worlds), grasses (grassland worlds), tree-grass coexistence, and bi-directionality with the system converging to grassland or to forest depending on the initial vegetation fractions. The propagation of seeds across latitudes leads to a widening of the coexistence region (Fig. \ref{fig:BauMaps}, right panel).\\

\begin{figure}
    \centering
    \includegraphics[width=68mm]{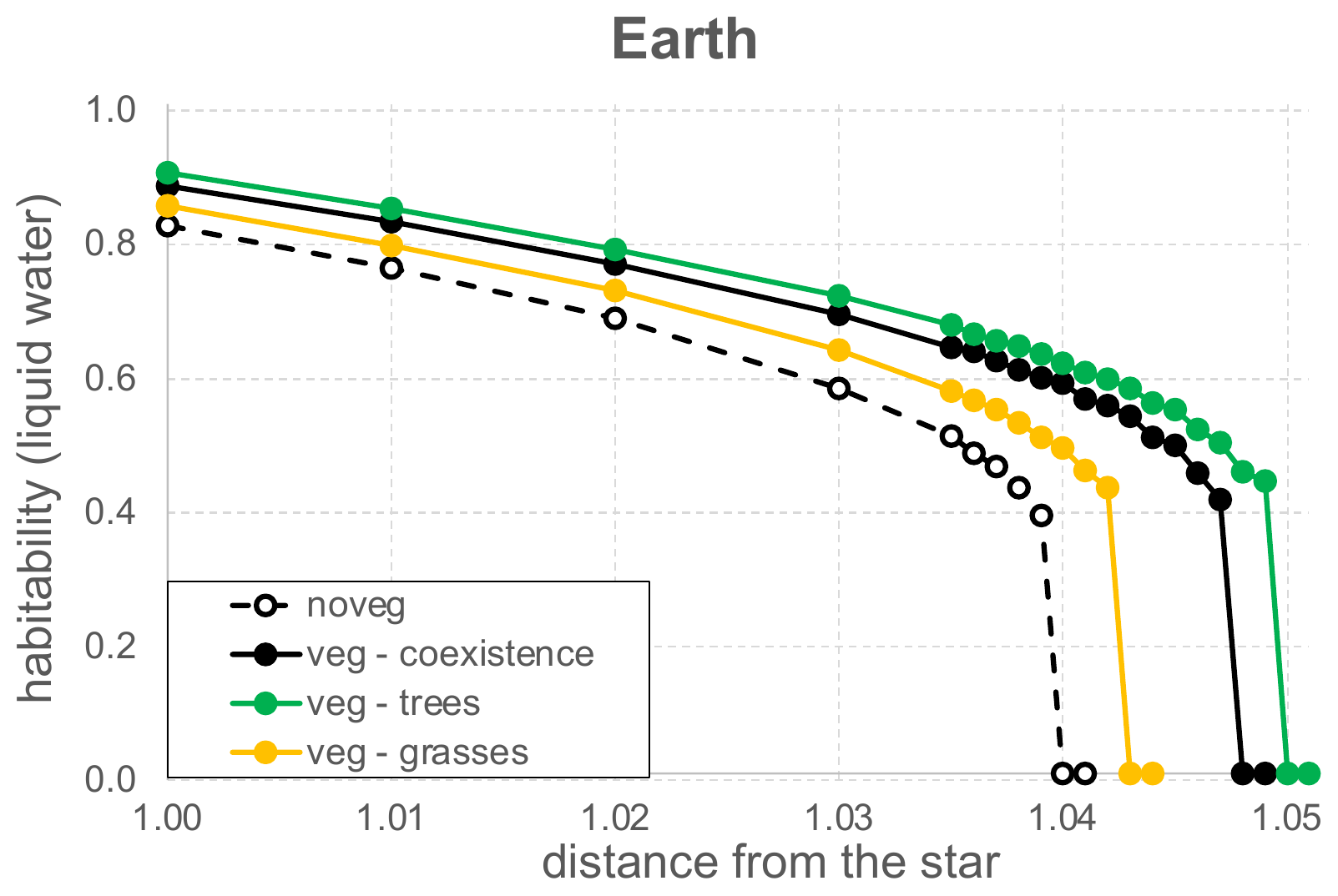}
    \caption{Impact of vegetation on the liquid water habitability index $h$ for the \textit{Earth}, as a function of the semi-major axis of the orbit $d$ (in this configuration, approximately conceivable as the distance from the star), for the cases of tree-dominance (continuous green/dark line), grass-dominance (continuous yellow/light line) and coexistence between trees and grasses (continuous black line). The dashed black line indicates the case without vegetation.}
    \label{fig:OuterEdgeEarth}
\end{figure}

\begin{figure}
    \centering
    \includegraphics[width=68mm]{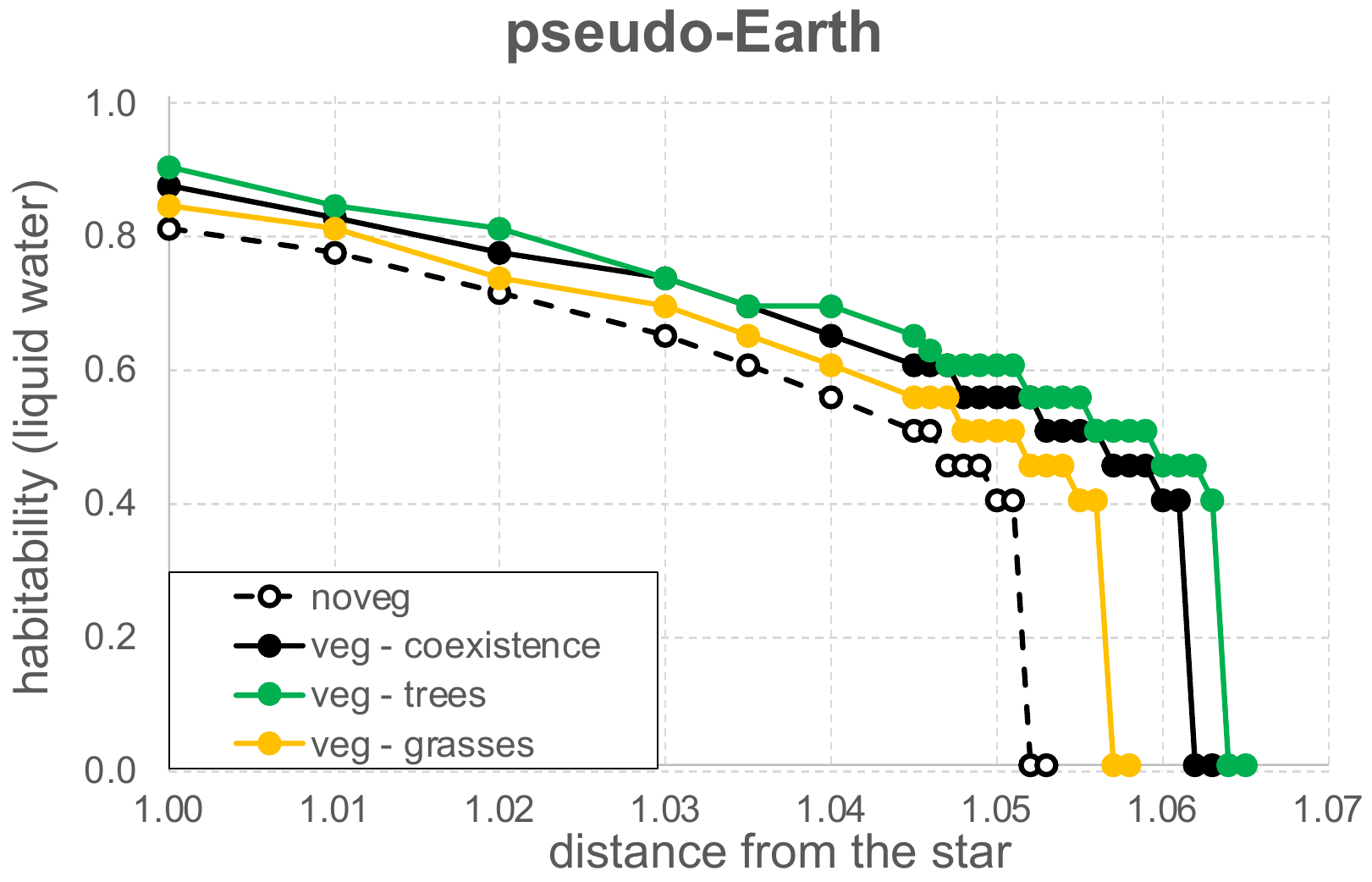}
    \caption{Impact of vegetation on the liquid water habitability index $h$ for the \textit{pseudo-Earth}, as a function of the semi-major axis of the orbit $d$ (in this configuration, strictly corresponding to the distance from the star), for the cases of tree-dominance (continuous green/dark line), grass-dominance (continuous yellow/light line) and coexistence between trees and grasses (continuous black line). The dashed black line indicates the case without vegetation.}
    \label{fig:OuterEdgePseudoEarth}
\end{figure}

\begin{figure}
    \centering
    \includegraphics[width=68mm]{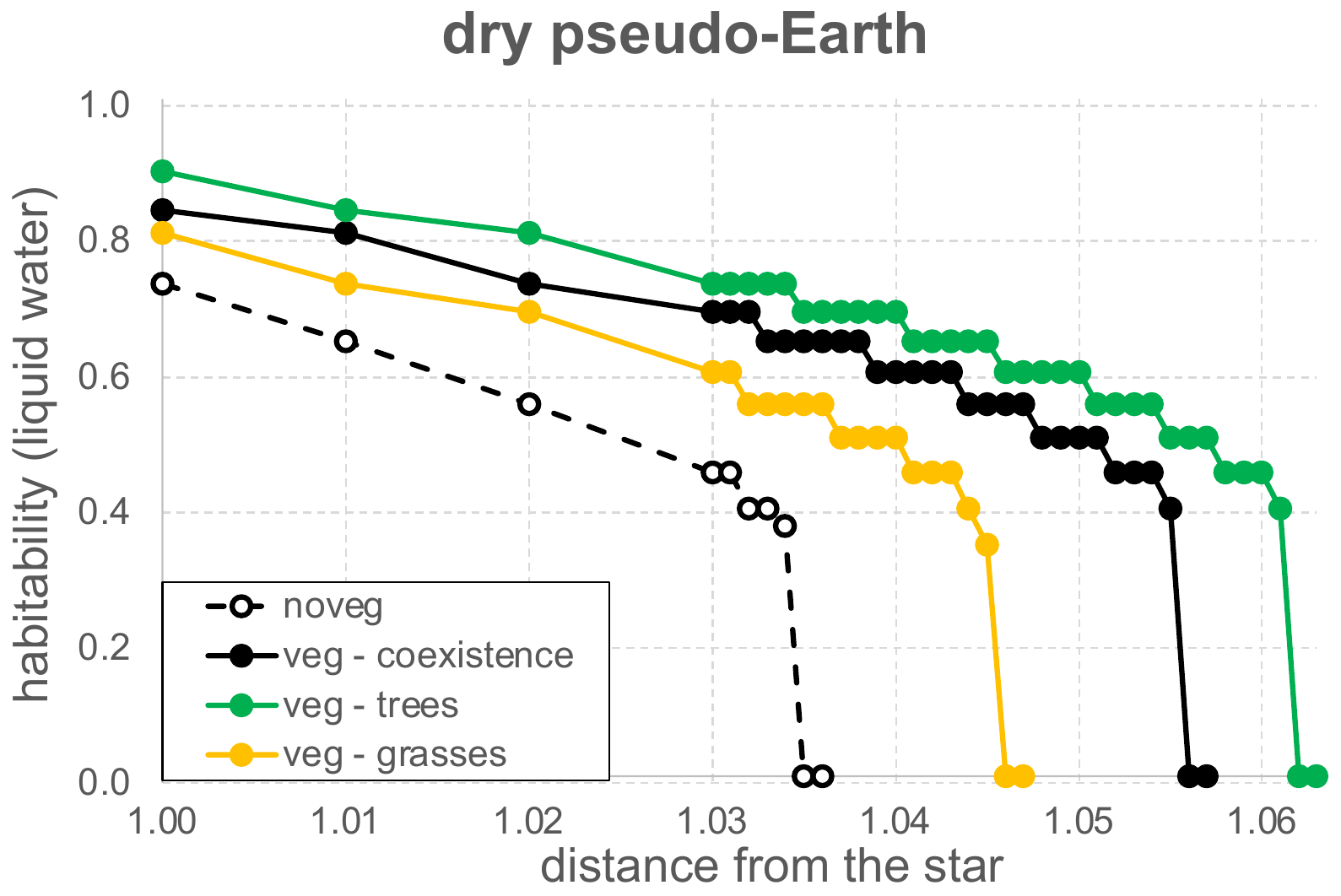}
    \caption{Impact of vegetation on the liquid water habitability index $h$ for the \textit{dry pseudo-Earth}, as a function of the semi-major axis of the orbit $d$ (in this configuration, strictly corresponding to the distance from the star), for the cases of tree-dominance (continuous green/dark line), grass-dominance (continuous yellow/light line) and coexistence between trees and grasses (continuous black line). The dashed black line indicates the case without vegetation.}
    \label{fig:OuterEdgePseudoDryEarth}
\end{figure}

\indent
Due to changes in the albedo caused by vegetation, the habitable zone and overall planetary habitability are extended beyond the traditional outer limits. (Figs. \ref{fig:OuterEdgeEarth}, \ref{fig:OuterEdgePseudoEarth}, \ref{fig:OuterEdgePseudoDryEarth}). 
Although the warming effect of vegetation due to the lowered albedo is to be expected, in our approach the vegetation distribution across latitudes is generated by the outcome of the dynamical competition for resources between trees and grasses, which is, in turn, determined by the average temperature in each latitudinal band. In this way, a full coupling between vegetation cover and temperature is simulated, based on non-linear feedback between these two variables. In general, thus, the achieved temperature-vegetation state is not imposed, but it \textit{emerges from} the dynamics of the vegetation-climate system.

\begin{figure}
    \centering
    \includegraphics[width=80mm]{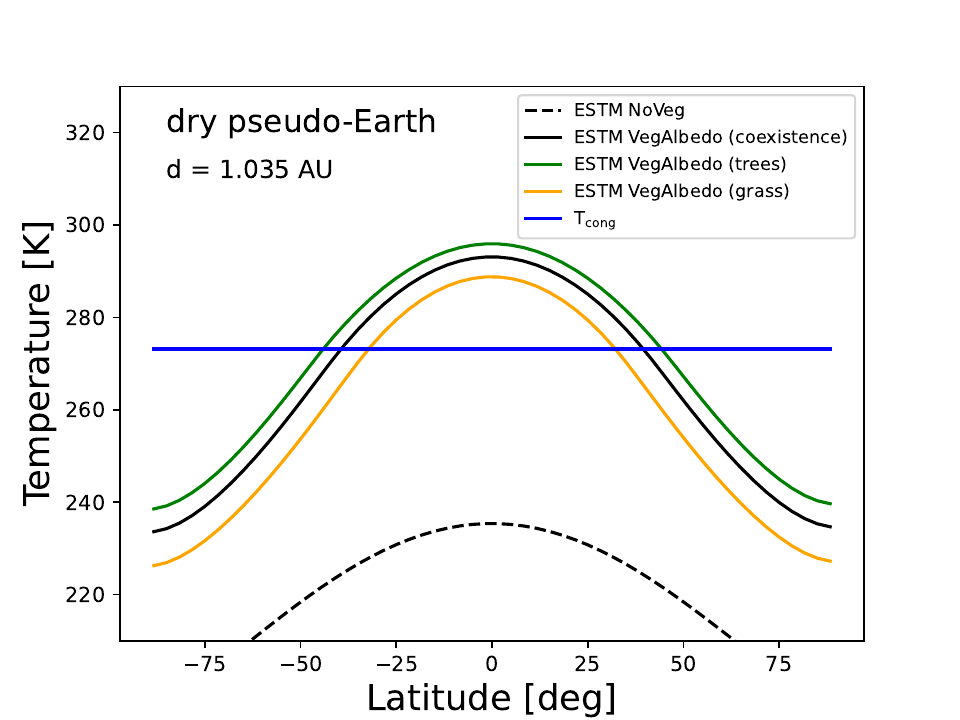}
    \caption{Latitudinal temperature profiles for the \textit{dry pseudo-Earth} at $d$ = 1.035 AU. Dashed line: case without vegetation. Continuous lines: cases with vegetation-albedo feedback. $T_\mathrm{cong}$: water freezing temperature.}
    \label{fig:OuterEdgePseudoDryEarthLat}
\end{figure}

This effect is especially relevant for planets that have a larger extension of continents than Earth and a correspondingly smaller ocean cover, such as the case we called \textit{dry pseudo-Earth}. In this case, in fact, the albedo change induced by vegetation has a stronger effect (cf. \citep{Abe2011}. However, the ocean fraction cannot be too small, as in this case the whole hydrological cycle could be modified. For example, on a vast continent, precipitation in the center of the continent could be severely limited. Therefore, if the continents were too large, their central parts could be unsuitable as habitats for vegetation. A complete quantification of these effects should include a full treatment of the planetary water cycle, which is beyond the scope of the present study.\\

\indent
Although relatively limited, the effect of vegetation albedo explored here indicates that the feedback of the biosphere on climate and, thus, on habitability itself cannot be discarded when analysing potential planetary habitability and the role of vegetation cover on continents.
Note that here we are considering a modern Earth-like chemical composition of the atmosphere, thus we do not include any attempt to model the maximum greenhouse limit for a CO$_\text{2}$-rich atmosphere as done, e.g., in \citep{Kasting1993} and \citep{Kopparapu2013}.
In our simulations, the outer edge of the circumstellar habitable zone is, in practice, the snowball limit for a planet with an Earth-like atmospheric composition
\footnote{Since we explored the impact of vegetation on the snowball edge of the habitable zone, the liquid-water habitability index adopted here, $h$, provides the same results of the “complex-life” index of habitability, $h_\mathrm{050}$, which is defined in the temperature range $0 \leq T  (^\circ\text{C})\leq 50$ \citep{Silva2017}.}.\\ 
\indent
An important consequence of the introduction of vegetation is therefore the passage from snowball conditions to a planet with intermediate habitability at the outer edge of the circumstellar habitable zone (Fig. \ref{fig:OuterEdgePseudoDryEarthLat}), which is especially important for habitable planets that are prone to snowball transition \citep{Murante2020}. 
We recall that, with respect to a world with bare granite continents, the effect of vegetation albedo is to increase the average surface temperature. On the other hand, grasses and trees have different albedo, so they affect temperature to differing degrees. In a world where both types of vegetation are present, such as the case considered here, the dominance of one type over another can lead to temperature changes and, in case the vegetation type with lower albedo is favoured at higher temperature, one can obtain a temperature-stabilising effect that is analogous to the homeostatic behavior of white and black daisies of Daisyworld \citep{WatsonLovelok1983}. This situation is especially interesting for planets in the grassland-forest coexistence regime, where small variations in external conditions (e.g., stellar luminosity or orbital variations) can be buffered by different latitudinal extensions of grasses and forests.\\

The dynamics explored here is extremely simplified and represents only a first step in the analysis of vegetation-habitability interactions. The adopted vegetation distribution is stripped at its bare minimum, and in reality the relative dominance of grasses and forests is related also to other parameters such as the availability of water. The vegetation types we considered represent boreal forests and low-latitude grasslands and savannas, and we did not include neither high-latitude grasslands and steppe, nor (more importantly) equatorial forests. In particular, grasses expanded in the last tens of million of years, when the Earth surface temperature decreased from the high values of the Eocene and the more arid conditions favoured grasses \citep{Cerling1993}. Usually,  the intensity of the hydrological cycle is considered to be roughly related to the average temperature \citep{Trenberth1998}, with colder climates having less precipitation than what happens in warmer conditions. Although the relation between temperature and precipitation is definitely much more complicated than this, we could anyway make the hypothesis that lower optimal temperatures are associated to grasses (favoured by the more arid conditions), while higher optimal temperatures are associated to trees. In this case, forests would dominate at lower latitudes and grasses/steppe would dominate at higher latitudes. Also in this case, the effect of vegetation is to warm up the planet with respect to bare soil. In coexistence regime, low-latitude trees have lower albedo than low-latitude grasses and, since most solar radiation impinges at low latitudes, the warming effect of vegetation is enhanced. Future work will explore in detail the effects of different choices of optimal temperatures and more complex vegetation configurations (e.g., trees at both low and high latitudes and grasslands at intermediate latitudes).\\

Another important question concerns the effect of vegetation on the atmospheric concentration of CO$_{\text{2}}$. Plants both sequester carbon from the atmosphere by photosynthesis, thus lowering the greenhouse effect, and emit CO$_{\text{2}}$ by respiration. To this, one should add the carbon dioxide emission by soil decomposition processes. In an equilibrium situation, the two competing effects of CO$_{\text{2}}$ removal and emission should balance each another, but temperature and CO$_{\text{2}}$ changes can strengthen one of the two components, and the net outcome is not easy to predict.
The time-scale of this process is the same of the vegetation life cycle (tens or hundreds of years); therefore, the feedback associated with the biological carbon cycle can be easily studied with ESTM. Future work will also include a simplified carbon balance model in the study of planetary habitability, possibly combined with parameterized descriptions of the much slower geological carbon dynamics associated with the carbonate-silicate cycle.

\section*{Acknowledgements}
This work is supported by the Italian Space Agency within the \textit{Life in Space} (ASI N. 2019-3-U.0) and \textit{ASTERIA} (ASI N. 2023-5-U.0) projects.\\ 
PS acknowledges further support by OGS and CINECA with the HPC-TRES program award N. 2022-02.\\
Part of this work has been supported by the INAF MINI-GRANT RSN2 \textit{ClimHAB-RBA} 1.05.12.04.02.\\
Computations have been performed using the clusters \texttt{amonra} \citep{Bertocco2020, Taffoni2020} and \texttt{pleiadi}, thanks to INAF computing time.\\
We also warmly thank the anonymous Referee for very useful comments that contributed to improve the quality of the manuscript.

\section*{Data Availability}
Data will be available from the authors upon request.
%The inclusion of a Data Availability Statement is a requirement for articles published in MNRAS. Data Availability Statements provide a standardised format for readers to understand the availability of data underlying the research results described in the article. The statement may refer to original data generated in the course of the study or to third-party data analysed in the article. The statement should describe and provide means of access, where possible, by linking to the data or providing the required accession numbers for the relevant databases or DOIs.



\bibliographystyle{mnras}
\bibliography{Bisesi-2024} % if your bibtex file is called example.bib

%\begin{thebibliography}{99}
%\end{thebibliography}

%%%%%%%%%%%%%%%%% APPENDICES %%%%%%%%%%%%%%%%%%%%%

\appendix
\section{Supplementary material}
%If you want to present additional material which would interrupt the flow of the main paper, it can be placed in an Appendix which appears after the list of references.

Movies with the complete set of profiles for all possible configurations in the ($c_\mathrm{g}$, $c_\mathrm{ts}$) parameter space are accessible at the links:

\vspace{0.1cm}
\indent \footnotesize \href{https://adlibitum.oats.inaf.it/bisesi/Bisesi-et-al_NoSeeds.mov}{https://adlibitum.oats.inaf.it/bisesi/Bisesi-et-al$\_$NoSeeds.mov}\\
\indent \footnotesize \href{https://adlibitum.oats.inaf.it/bisesi/Bisesi-et-al_NoSeeds-Inverted.mov}{https://adlibitum.oats.inaf.it/bisesi/Bisesi-et-al$\_$NoSeeds-Inverted.mov}\\
\indent \footnotesize \href{https://adlibitum.oats.inaf.it/bisesi/Bisesi-et-al_Seeds.mov}{https://adlibitum.oats.inaf.it/bisesi/Bisesi-et-al$\_$Seeds.mov}\\
\indent \footnotesize \href{https://adlibitum.oats.inaf.it/bisesi/Bisesi-et-al_Seeds-Inverted.mov}{https://adlibitum.oats.inaf.it/bisesi/Bisesi-et-al$\_$Seeds-Inverted.mov}\\

%%%%%%%%%%%%%%%%%%%%%%%%%%%%%%%%%%%%%%%%%%%%%%%%%%

% Don't change these lines
\bsp	% typesetting comment
\label{lastpage}
\end{document}